\documentstyle[prl,aps,amsfonts,amssymb,floats,epsfig,graphicx]{revtex}

\begin{document}
\draft
\twocolumn[\hsize\textwidth\columnwidth\hsize\csname
@twocolumnfalse\endcsname
\title{Zero-field incommensurate spin-Peierls phase with interchain frustration in TiOCl}
\author{
R. R\"{u}ckamp,$^{1}$
J. Baier,$^{1}$ M. Kriener,$^{1}$
M.W. Haverkort,$^{1}$
T. Lorenz,$^{1}$ G.S. Uhrig,$^{2}$
\\
L. Jongen,$^{3}$ A. M\"{o}ller,$^{3}$ G. Meyer,$^{3}$
and M. Gr\"{u}ninger$^{4}$}
\address{
$^1$\it II. Physikalisches Institut, Universit\"{a}t zu K\"{o}ln,
50937 K\"{o}ln, Germany
\\
{\rm $^2\!$} Institut f\"{u}r Theoretische Physik, Universit\"{a}t zu K\"{o}ln,
50937 K\"{o}ln, Germany
\\
{\rm $^3\!$} Institut f\"{u}r Anorganische Chemie, Universit\"{a}t
zu K\"{o}ln, 50937 K\"{o}ln, Germany
\\
{\rm $^4\!$} II. Physikalisches Institut, RWTH Aachen,
52056 Aachen, Germany
}
\date{March 16, 2005}
\maketitle
\begin{abstract}
We report on the magnetic, thermodynamic and optical properties of the quasi-one-dimensional quantum antiferromagnets
TiOCl and TiOBr, which have been discussed as spin-Peierls compounds. The observed deviations from canonical
spin-Peierls behavior, e.g.\ the existence of two distinct phase transitions,
have been attributed  previously to strong orbital fluctuations.
This can be ruled out by our optical data of the orbital excitations.
We show that the frustration of the interchain interactions in the bilayer structure gives rise to incommensurate order
with a subsequent lock-in transition to a commensurate dimerized state.
In this way, a single driving force, the spin-Peierls mechanism, induces two separate transitions.
\end{abstract}
\vskip2pc]
%
\narrowtext


Low-dimensional quantum spin systems exhibit a multitude of interesting phenomena.
For instance a one-dimensional (1D) $S$=1/2 chain coupled to the lattice may show
a spin-Peierls transition to a non-magnetic, dimerized ground state. In recent years,
detailed studies of the first inorganic spin-Peierls compound CuGeO$_3$ have deepened
the understanding of spin-Peierls systems substantially \cite{lemmens}.
Even richer physics is expected if the spins are coupled additionally to orbital or
charge degrees of freedom.
A prominent example is the complex behavior of NaV$_2$O$_5$, which arises from the
interplay of spin dimerization, orbital order and charge order \cite{lemmens}.
Recently, TiOCl and TiOBr have been discussed as spin-Peierls compounds with strong
orbital fluctuations
\cite{seidel03,kataev03,imai,lemmenstiocl,caimi03,caimi04,sahadasgupta,hemberger},
assuming a near degeneracy of the $t_{2g}$ orbitals in these $3d^1$ systems.
Different quantities such as the magnetic susceptibility \cite{seidel03}, the specific
heat \cite{hemberger}, ESR data \cite{kataev03} and NMR spectra \cite{imai} point towards
the existence of two successive phase transitions, which clearly goes beyond a canonical
spin-Peierls scenario in which a single second-order phase transition is expected.
The high transition temperatures of $T_{c1}$=67\,K and $T_{c2}$=91\,K found in TiOCl are
fascinating in a spin-Peierls context.

The structure of TiOX consists of 2D Ti-O bilayers within the $ab$ plane which are well separated
by X=Cl/Br ions \cite{smaalen}. Quasi-1D $S$=1/2 chains are formed due to the occupation of the $d_{y^2-z^2}$
orbital in the ground state (see below), giving rise to strong direct exchange between
neighboring Ti sites along the $b$ axis ($y$ direction) and negligible coupling in the other
directions. Accordingly, the magnetic susceptibility of TiOCl is well described at high temperatures
by the 1D $S$=1/2 Heisenberg model with an exchange constant of $J\approx $~676\,K \cite{seidel03,kataev03}.
In the non-magnetic low-temperature phase, a doubling of the unit cell along the chain direction
has been observed by x-ray measurements for both TiOCl \cite{smaalen} and TiOBr \cite{sasaki},
supporting a spin-Peierls scenario.
However, the following experimental facts are not expected in a canonical spin-Peierls system:
(i) the existence of two successive phase transitions \cite{seidel03,kataev03,imai,hemberger},
(ii) the first-order character of the low-temperature phase transition \cite{hemberger,smaalen,sasaki},
(iii) the observation of inequivalent Ti sites by NMR at low temperatures \cite{imai}, and
(iv) the appearance of incommensurate superstructure satellites between $T_{c1}$ and $T_{c2}$
\cite{abel,smaalenpriv}, which arise in a generic spin-Peierls system only in a sufficiently
high magnetic field.
Testing whether TiOX displays the generic $H$-$T$ phase diagram of a spin-Peierls system \cite{cross}
is difficult due to the large energy scale.

The aim of the present study is to determine the ingredients necessary to understand the
peculiar properties of TiOX.\@
Using optical spectroscopy, we show that the $t_{2g}$ subshell on the Ti sites experiences a
pronounced crystal-field splitting of 0.65\,eV.\@ Thus the orbital degree of freedom is quenched
and can be neglected for the description of the low-energy physics.
All of the above mentioned points can be attributed to the coupling of spin and lattice degrees of
freedom in the bilayer structure, which gives rise to a frustration of interchain interactions.


Single crystals of TiOCl and TiOBr have been grown by the chemical vapor transport technique \cite{schaefer}.
The purity of the crystals was checked by x-ray powder diffraction.
Typical crystal dimensions are a few mm$^2$ in the $ab$ planes and 10--100~$\mu$m along the $c$
axis, i.e., the stacking direction.
For the measurements of the magnetic susceptibility $\chi$ and the specific heat $C_p$ we have employed a
vibrating sample magnetometer, a SQUID magnetometer, a Faraday balance and different calorimeters using
the relaxation time and continuous heating method.
In order to obtain a large enough signal we have co-aligned several crystals (with the $c$ axis parallel to
the applied magnetic field of 1~Tesla) for the measurements of $\chi$.
For the determination of $C_p$ we pressed pellets from crashed single crystals of about 28.8~mg.
The linear thermal expansion $\alpha = \partial \ln L\,/\,\partial T$ has been measured on 7 single crystals
stacked on top of each other with $L||c$ by a home-built capacitance dilatometer.
Optical data have been collected on thin single crystals using a Fourier spectrometer.

\begin{figure}[t]
\centerline{\psfig{figure=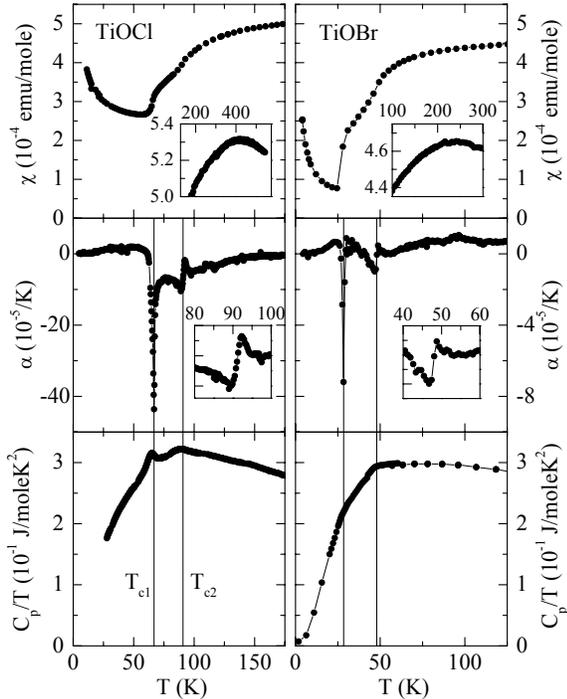,width=7.5cm,clip=}}
  \caption{
  Magnetic susceptibility $\chi$, thermal expansion $\alpha$, and specific heat $C_p$
  of TiOCl (left panels) and TiOBr (right).}
\label{thermodyn}
\end{figure}


In Fig.\ \ref{thermodyn} we plot $\chi$, $\alpha$ and $C_p/T$ of TiOCl and TiOBr. The existence of two distinct
phase transitions at $T_{c1}$ and $T_{c2}$ (67 and 91\,K in TiOCl; 28 and 48\,K in TiOBr) is most clearly detected in $\alpha$,
but anomalies are also observed in $\chi$ and $C_p/T$ at both $T_{c1}$ and $T_{c2}$. The very similar behavior
of TiOCl and TiOBr proves that the occurrence of {\em two} transitions is an intrinsic property
of these compounds. The sharp and symmetric anomaly of $\alpha$ at $T_{c1}$ signals an almost jump-like decrease of the lattice
parameter $c$ with increasing temperature indicating a first-order transition, in agreement with an analysis
of $C_p$ of TiOCl \cite{hemberger} and x-ray data of the superstructure satellites
\cite{smaalen,sasaki}.
The anomaly of $\alpha$ at $T_{c2}$ does not display the typical shape of a phase transition, neither of first nor of
second order, but rather resembles a glass transition \cite{muellerlang}.
Recently, the appearance of incommensurate superstructure satellites below $T_{c2}$ has been reported \cite{abel,smaalenpriv},
which may be reconciled with a transition with glass-like features due to pinning to impurities.
Below $T_{c1}$, $\chi$ is dominated by the van Vleck term and by contributions of impurities and of remnants
of organic solvents \cite{kataev03}.
Above $T_{c2}$, $\chi$ can be described by a 1D $S$=1/2 Heisenberg chain \cite{kluemper}
with exchange constant $J\approx$~676\,K in TiOCl~\cite{seidel03,kataev03}.
For TiOBr the position $T_{\rm max}\! \approx \!$~240\,K of the maximum of $\chi$ allows to estimate $J\! \approx \!$~375\,K
from $T_{\rm max}/J \approx 0.64$ \cite{kluemper}, but $\chi(T)$ deviates from the theoretical curve \cite{sasaki,lemmens05}.
The weak anomalies of $C_p$ show that only a small amount of entropy is released
at the phase transitions \cite{hemberger}, in agreement with the expectations for a spin-Peierls transition of a
1D $S$=1/2 chain, where most of the magnetic entropy $R \ln 2$ is connected with short-range correlations
which are present up to temperatures of order $J \gg T_{c1}$, $T_{c2}$.


\begin{figure}[t]
\centerline{\psfig{figure=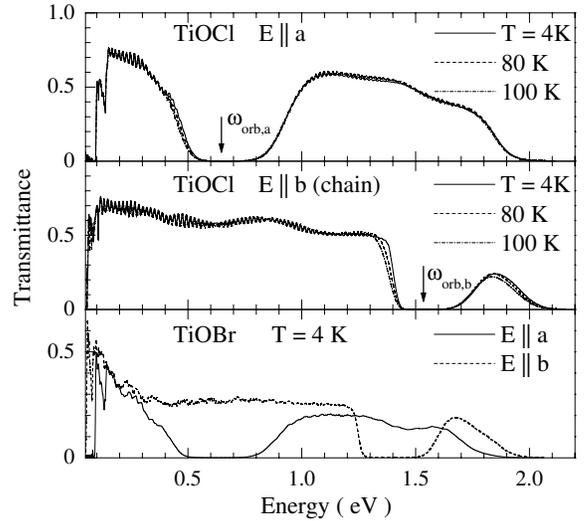,width=7.5cm,clip=}}
  \caption{Transmittance $T(\omega)$ of TiOX.
  For large values of $T(\omega)$, the spectra are dominated by interference fringes.
}
\label{MIRtrans}

\end{figure}

In the literature, the peculiar properties of TiOX have been attributed tentatively to
strong orbital fluctuations
\cite{seidel03,kataev03,imai,lemmenstiocl,caimi03,caimi04,sahadasgupta,hemberger}.
We have determined the orbital excitation energies, i.e., the crystal-field splitting
of the $3d$ levels, by measuring the transmittance of thin single crystals (Fig.\ \ref{MIRtrans}).
The transmittance is suppressed below $\approx 0.1$\,eV by phonon absorption and above
$\approx 2$\,eV by interband excitations, i.e.,
excitations across the charge gap. The absorption at 0.6-0.7\,eV for $E\! \parallel \! a$
and at 1.3-1.6\,eV for $E \! \parallel \! b$ can be identified as orbital excitations which are
infrared active due to the lack of inversion symmetry on the Ti sites.
Both the excitation energies and the polarization dependence are in good agreement with the
predictions of a cluster calculation of the crystal-field levels \cite{njop}.
The observed features correspond to transitions from the ground state (predominantly $d_{y^2\!-\!z^2}$
with a small admixture of $p_z$ character) to the second and third excited states. Transitions to
the first excited state, $d_{xy}$, are not directly infrared active in the 300\,K structure.
Our cluster calculation \cite{njop} indicates a value of $\approx \!$~0.3\,eV ($\approx \! $~3500\,K)
for the $d_{xy}$ level. Moreover, we can exclude a significant admixture
of the $xy$ orbital to the ground state since a population of the $xy$ orbital would lead to
an entirely different polarization dependence \cite{njop}. Finally, we observe only a small change
of the respective absorption features as a function of temperature (see Fig.\ \ref{MIRtrans}).
We conclude that orbital fluctuations are strongly suppressed and do not constitute the origin
of the peculiar properties of TiOX.\@
This is in agreement with recent LDA+U and LDA+DMFT results \cite{sahadasgupta04b,pisani}
and with an analysis of ESR data \cite{kataev03}.

\begin{figure}[t]
\centerline{\psfig{figure=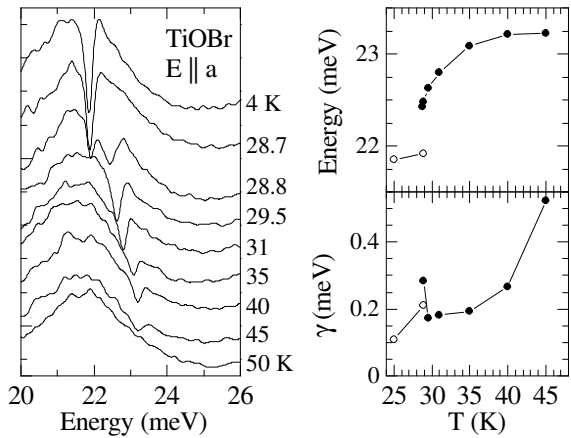,width=7.5cm,clip=}}
  \caption{Left: far-infrared transmittance of a thin single crystal of TiOBr
  for $E \parallel a$. The broad sinusoidal feature corresponds to a Fabry-Perot interference
  fringe. The sharp dip shifting from 23 to 22 meV reflects absorption by a phonon which
  becomes infrared active due to structural changes at $T_{c2}$.
  Right: temperature dependence of the phonon energy and of the line width $\gamma$.
}
\label{newmodes}
\end{figure}

With a charge gap of $\approx \!$~2\,eV and an orbital excitation gap of the order of 0.3\,eV,
it is sufficient to consider spin and lattice degrees of freedom for the explanation
of the low-energy physics.
Clear evidence for structural changes at both $T_{c1}$ and $T_{c2}$ is provided by an
analysis of the phonon spectra. We have observed changes of the number of infrared-active
phonons in far-infrared reflectance and transmittance measurements on TiOCl and TiOBr
at both $T_{c1}$ and $T_{c2}$. As an example we show the temperature dependence of a
phonon observed for $E\! \parallel \! a$ in TiOBr in Fig.\ \ref{newmodes}.
At 4\,K, the energy of this mode amounts
to 21.8\,meV in TiOBr and 22.3\,meV in TiOCl. This small difference shows that the Br/Cl ions
hardly contribute to this mode.
The energy, polarization and temperature dependence indicate a displacement of mainly Ti ions along
the $a$ axis, with neighboring Ti ions within a chain moving out-of-phase.
In the dimerized superlattice structure this mode is folded back to the Brillouin zone center.
It becomes weakly infrared active if neighboring Ti sites are inequivalent, as observed by
NMR \cite{imai}.
The left panel of Fig.\ \ref{newmodes} shows that this mode gains finite spectral weight at about
$T_{c2}$=48\,K in TiOBr. The line width is reduced by a factor of three upon cooling from 45 to 29\,K,
and also the peak energy is strongly temperature dependent, showing a jump at $T_{c1}$.
These observations are in agreement with an incommensurate intermediate phase \cite{abel,smaalenpriv},
which locally resembles the low temperature phase.

\begin{figure}[t]
\centerline{\psfig{figure=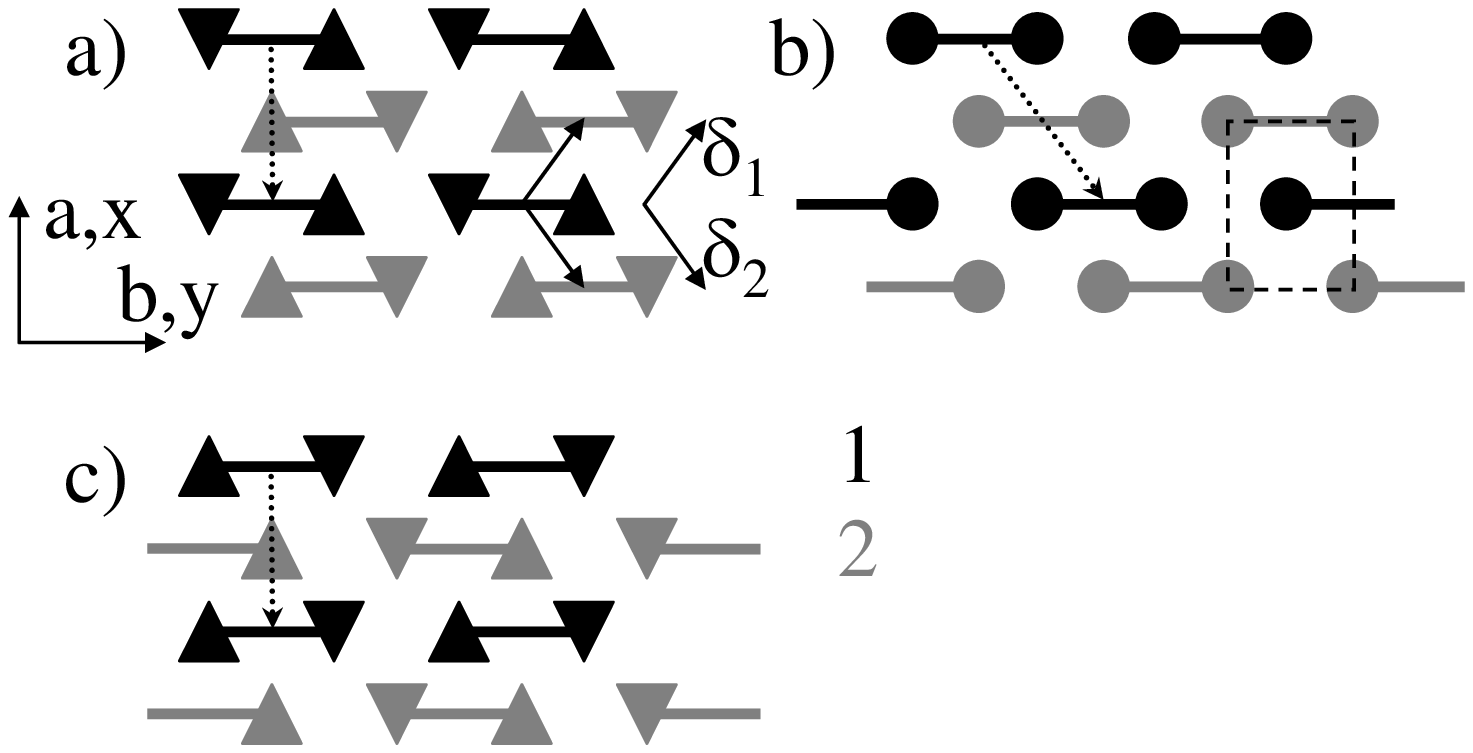,width=7.5cm,clip=}}
  \caption{Patterns of Ti spin dimers in the bilayer structure
  of TiOX.\@
  Black (gray): Ti sites in layer 1 (2) of a single bilayer.
  (a) in-phase arrangement of neighboring chains within a layer.
  (b) out-of-phase arrangement (see dashed arrows).
  The choice of $\delta_1$ and $\delta_2$ is convenient for the theoretical description.
  }
\label{bilayer}
\end{figure}


We now turn to the bilayer geometry and discuss different scenarios for the existence of two
phase transitions. In a single $S$=1/2 chain, the dimers may reside either on the even or
on the odd bonds. In an array of chains, the dimerization on nearest-neighbor chains may be
in-phase or out-of-phase with respect to each other.
In the bilayer structure of TiOX, a Ti ion of layer 1 is located on top of the center of a plaquette
of four Ti ions on layer 2 (dashed box in Fig.\ \ref{bilayer}b). First we discuss in-phase and
out-of-phase dimer patterns within the individual layers.
In the out-of-phase case, all Ti sites are equivalent (circles in Fig.\ \ref{bilayer}b), each
facing one dimer on the surrounding plaquette. However, there are two inequivalent Ti sites in the
in-phase case, facing either two or zero dimers (up/down triangles in Fig.\ \ref{bilayer}a).
The observation of two distinct Ti sites in NMR \cite{imai} at low temperatures thus clearly indicates
that an in-phase pattern is realized.
This is corroborated by structural data \cite{smaalen}, which show that two Ti ions forming a
dimer are displaced with opposite signs perpendicular to the layers.

For a single bilayer there are four degenerate in-phase patterns, with dimers on the even/even,
even/odd, odd/even or odd/odd bonds of layers 1/2 (see Fig.\ \ref{bilayer}a). In principle, this
allows for several phase transitions between different superstructures, e.g., from
an undimerized high-temperature phase to a dimerized phase where all bilayers realize the even/even
pattern and finally to a low-temperature phase in which the dimerization alternates between even/even
and odd/odd on adjacent bilayers. However, in this scenario both the NMR signal of two inequivalent Ti
sites and the commensurate superstructure satellites are expected to appear at the high-temperature
phase transition at $T_{c2}$, in contrast to the experimental observation.

Our second scenario focuses on the observation of two inequivalent Ti sites at low
temperatures \cite{imai,smaalen}. In principle, this inequivalence may serve as a second order
parameter, i.e., the system may undergo both a spontaneous spin-Peierls transition and, at a different
temperature, a spontaneous transition to a state with Ti site inequivalence.
In the present bilayer structure these two order parameters are coupled, but a Landau expansion shows
that there still may be two distinct phase transitions. Roughly, this scenario predicts the onset of
dimerization with some admixture of site inequivalence at $T_{c2}$ and the development of full site
inequivalence below $T_{c1}$ (or vice versa).
In this scenario, commensurate superstructure satellites are predicted to appear at the high-temperature
phase transition at $T_{c2}$, in disagreement with the experimental data on both  TiOCl and
TiOBr \cite{imai,smaalen,sasaki}.

In the intermediate phase, superstructure satellites have been reported which are incommensurate both
in $b$ and $a$ direction \cite{abel,smaalenpriv}. Incommensurate order was also proposed in
order to explain the very broad NMR signal in the intermediate phase \cite{imai}.
In compounds with commensurate band filling, incommensurability may arise from the frustration of competing
interactions which favor different ordering wave vectors.
We propose that the incommensurability arises from the frustration inherent to the bilayer structure.
The Landau expansion for the free energy as a function of the displacement $\phi_{i}^y$ of the Ti ion $i$
parallel to the chains reads to quadratic order
\begin{eqnarray}
\Delta F & = &
  \frac{a_0}{2}\sum_i (\phi_{i}^y)^2 + \frac{a_1}{2}\sum_i \phi_{i}^y\phi_{i+\delta_1 +\delta_2}^y
\\
\nonumber
& & + \frac{b}{2}\sum_i \phi_{i}^y (\phi_{i+\delta_1}^y + \phi_{i-\delta_2}^y)
\\
\nonumber
 = & \frac{1}{2} & \sum_{(h,k)} |\phi_{(h,k)}^y|^2(a_0 + a_1 \cos(k) + 2b \cos(\frac{h}{2})\cos(\frac{k}{2}))
\end{eqnarray}
where $h$ and $k$ denote the momenta perpendicular and parallel to the chains, respectively.
See Fig.\ \ref{bilayer}a for the definition of $\delta_1$ and $\delta_2$.
The $a_1$ term describes the tendency towards a spin-Peierls distortion ($a_1 \! > \! 0$).
For vanishing interlayer coupling, i.e.\ $b$=0, the system undergoes spontaneous dimerization with
$k\!=\!\pi$  if $a_0\!-\!a_1\! <\! 0$.
However, for $b \neq 0$ the free energy is minimized for $h$=0 and $k\!=\!2 \arccos(-b/2a_1) \approx \pi+b/a_1$,
i.e., the system becomes incommensurate for any finite value of the interlayer coupling $b$.
This is due to the fact that the interlayer coupling described by the $b$ term vanishes for $k\!=\!\pi$.
The system has to become incommensurate in order to gain energy
from the interlayer coupling.

In order to explain the additionally observed incommensurability perpendicular to the chains we have to consider
a coupling of $\phi_i^y$ to the displacements $\phi_j^x$ and $\phi_j^z$ in $a$ and in $c$ direction, where $i$
and $j$ are neighbors on adjacent chains as described for the $b$ term  above.
The formation of a dimer on sites $i$ and $i\!+\!\delta_1 \!+\!\delta_2$ and the corresponding $\phi_i^y$
and $\phi_{i+\delta_1+\delta_2}^y$ push away the Ti ion on site $i\!+\!\delta_1$, i.e., they couple to $\phi_{i+\delta_1}^{x,z}$.
This gives rise to an effective {\em intralayer} coupling between the chains, i.e., between site $i$ and
$i\! + \! \delta_1 \! - \! \delta_2$.
The coupling via $\phi^z$ leads to a term approximately proportional to $-\cos(h/2)$, favoring the in-phase pattern
(Fig.\ \ref{bilayer}a). In contrast, the coupling via $\phi^x$ produces a term approximately proportional to $\sin(h/2)$,
favoring the out-of-phase pattern. In total this yields an incommensurate value of $h$.

Experimental evidence for a coupling between $\phi^y$ and $\phi^{x,z}$ stems from the observation of finite values of
$\phi^z$ in the distorted low-temperature phase showing the in-phase pattern \cite{smaalen}, and from the strong temperature
dependence of the phonon observed for $E \! \parallel \! a$ depicted in Fig.\ \ref{newmodes}.
Furthermore, this scenario has two consequences which can be tested experimentally. First, it predicts finite values of
$\phi_i^x$ in the intermediate phase. Second, the incommensurability $\Delta k \approx b/a_1$ is predicted to decrease
with decreasing temperature, because the tendency $a_1$ towards a spin-Peierls distortion grows with decreasing
temperature. For low temperatures and small enough $\Delta k$, a first-order lock-in transition to commensurate
order occurs, as observed.

In conclusion, we have shown that TiOCl and TiOBr do not display canonical spin-Peierls behavior.
The peculiar properties of TiOX cannot be attributed to orbital fluctuations due to the large
crystal-field splitting of the $3d$ levels. The bilayer structure offers a clear explanation
for the existence of two distinct phase transitions.
The incommensurate phase at intermediate temperatures arises due to the frustration of interchain
interactions.
These compounds offer the possibility to study a spin-Peierls transition in a predominantly two-dimensional,
frustrated lattice.

It is a pleasure to acknowledge fruitful discussions with D.I. Khomskii, R. Valenti,
S. Scheidl, D. Fausti, P. van Loosdrecht, M. Mostovoy, S.-L. Drechsler and M. Lang.
This project is supported by the DFG via SFB 608.


\end{document}